\documentclass[11pt,a4paper,notitlepage,fleqn]{article}

\usepackage{setspace}
\usepackage[margin=1in]{geometry}
\usepackage[symbol]{footmisc}

\usepackage[T1]{fontenc}
\usepackage[utf8]{inputenc}
\usepackage{authblk}

\makeatletter

\newcommand{\fboxsubsec}[1]{
	\begin{flushleft}
		#1
	\end{flushleft}
	}
\renewcommand{\subsection}{\@startsection{subsection}{2}{0pt}
	{1ex}
	{0.5ex}
	{\reset@font\it\fboxsubsec}
	}
\makeatother

\title{A Note on the Asymptotic Properties of the GLS Estimator in Multivariate Regression with Heteroskedastic and Autocorrelated Errors}

\author{Koichiro Moriya$^{a}$\thanks{\scriptsize E-mail: moriya.koichiro@keio.jp (Corresponding Author), Tel. +81-466-49-3404.} \ and \ Akihiko Noda$^{b}$\thanks{\scriptsize E-mail: anoda@meiji.ac.jp, Tel. +81-3-3296-2265.}

{\scriptsize ${}^{a}$ \it Graduate School of Media and Governance, Keio University, 5322 Endo, Fujisawa, Kanagawa 252-0882, Japan} 

{\scriptsize ${}^{b}$ \it School of Commerce, Meiji University, 1-1 Kanda-Surugadai, Chiyoda-ku, Tokyo 101-8301, Japan}}

\date{\empty}

\vspace{5mm}

\renewcommand\thefootnote{\arabic{footnote}}

\usepackage{graphicx}
\usepackage[sort]{natbib}%
\usepackage{amsmath,amsthm,amssymb}
\usepackage{ascmac}
\usepackage{multirow}%
\usepackage{lscape}%
\usepackage{subfigmat}

\usepackage{pifont}%
\usepackage{arydshln}%
\usepackage[format=hang]{caption}
\usepackage[all]{xy}

\PassOptionsToPackage{hyphens}{url}
\usepackage{color}
\usepackage[dvipdfmx,
            bookmarkstype=toc,
	    colorlinks=true, 
	    pdfborder={0 0 1}, 
	    bookmarks=true, 
	    bookmarksnumbered=true, 
	    citecolor=blue]{hyperref}

\bibpunct{(}{)}{;}{a}{}{,}
\usepackage{here}
\usepackage{booktabs}
\usepackage{mathrsfs}
\usepackage{mathtools}
\def\hsymbu#1{\smash{\lower1.7ex\hbox{\huge$#1$}}}

\def\ve #1{{\mbox{\boldmath $#1$}}}

\DeclareMathOperator*{\vecop}{vec}

\DeclareMathOperator{\rank}{rank}

\newtheorem{theorem}{Theorem}

\newtheorem{assumption}{Assumption}

\newtheorem{corollary}{Corollary}

\newcommand{\bm}[1]{\mbox{\boldmath{$#1$}}}

\newcommand{\citetapos}[1]{\citeauthor{#1}'s \citeyearpar{#1}}
\newcommand{\citeapos}[2]{\citeauthor{#1}'s (\citeyear{#2})}
\newcommand{\ex}{{\mathbb{E}}}
\newcommand{\var}{{\rm Var}}
\newcommand{\cov}{{\rm Cov}}

\def\ve #1{{\mbox{\boldmath $#1$}}}

\begin{document}

\begin{titlepage}

\renewcommand{\thepage}{}
\renewcommand{\thefootnote}{\fnsymbol{footnote}}

\maketitle

\vspace{-10mm}

%\begin{quote}
\noindent
\hrulefill

\noindent
{\bfseries Abstract:} We study the asymptotic properties of the GLS estimator in multivariate regression with heteroskedastic and autocorrelated errors. We derive Wald statistics for linear restrictions and assess their performance. The statistics remains robust to heteroskedasticity and autocorrelation.\\



\noindent
\hrulefill
%\end{quote}

\end{titlepage}

\bibliographystyle{asa}

%\pagebreak

\section{Introduction}\label{sec:fgls_grs_intro}
Economists have extensively studied hypothesis testing procedures within regression frameworks to assess the validity of theoretical models. Typically, these statistical tests investigate whether certain necessary conditions hold under the assumption that the model is correctly specified. Specifically, several statistical tests rely on the theoretical fact that if an asset pricing model accurately explains the observed variation in stock returns, then the constant term in the model must be zero. A prominent example is the test proposed by \citet{gibbons1989teg}, commonly known as the GRS test, which examines the null hypothesis that the asset pricing model holds under the constraint that the constant term vector is zero. In the GRS test, the error terms are assumed to be independently and identically distributed according to a joint normal distribution. Despite these restrictive assumptions, the GRS test has been widely used in evaluating asset pricing models due to its clear and intuitive economic interpretation.

Some previous studies have raised concerns regarding the assumptions underlying the GRS test. For instance, \citet{affleck1989nta} examine the performance of the GRS test when the assumption of normality for the error terms is violated. They suggest that substantial departures from normality can distort both the size and power properties of the GRS test. Similarly, \citet{zhou1993apt} investigates how deviations from normality affect the size of the GRS test and demonstrates that such deviations lead to an increase in the Type I error rate. More recently, \citet{kamstra2024tra} propose a modified version of the GRS test to address finite-sample size distortions. However, their method does not address situations where the error terms deviate from multivariate normality.

To address situations in which error terms deviate from the normal distribution, \citet{kiefer2000srt} and \citet{kiefer2002hars} propose a heteroskedasticity and autocorrelation-robust (HAR) Wald test. However, the HAR Wald test suffers from the curse of dimensionality in nonparametric method due to employ \citetapos{newey1987sps} kernel-based estimator, potentially weakening finite-sample performance. In response to this issue, alternative kernel-based estimators is proposed in the literature such as \citet{phillips2007lrv} and \citet{ray2008pha}. Nonetheless, these alternative method still require a choice of hyperparameters, leading to potential variability in their performance.

In this paper, we study the asymptotic properties of the GLS estimator for the multivarite parametric regression models and derive Wald statistics based on the Prais-Winston (PW) and Cochrane-Orcutt (CO) type feasible GLS estimators. These Wald statistics are explicitly developed to accommodate heteroskedasticity and autocorrelation in error terms. In simulation experiments, we confirm that our proposed Wald tests perform well regardless of the presence of heteroskedasticity and autocorrelation. In contrast, the original and modified GRS tests exhibit severe Type I errors when error terms are autocorrelated, and the HAR Wald test tends to over-reject the null hypothesis as the number of cross-sectional units increases.

The rest of this paper is organized as follows. In Section \ref{sec:fgls_grs_model}, we show asymptotic properties of the GLS wstimator for multivariate regression model with heteroskedastic and autocorrelated errors, and derive the Wald statistics based on these estimators. In Section \ref{sec:fgls_grs_sim}, we provide simulation experiments to confirm the finite sample properties of our proposed Wald tests and the conventional tests. In Section \ref{sec:fgls_grs_con}, we conclude this study.

\section{Model and Test}\label{sec:fgls_grs_model}
In this section, we analyze the asymptotic properties of the GLS estimator in multiple regression with heteroskedastic and autocorrelated errors. In particular, we derive the asymptotic distribution for the PW type feasible GLS estimator and the Wald test statistics based on the estimator.

\subsection{The Model}\label{subsec:SU}
We consider the following model:
\begin{equation}
Y_{i,t} = \alpha_i + \mathbf{x}_{i,t}^{\prime}\bm{\beta}_i + e_{i,t},\ \ i=1,...,N,\ t=1,...,T,\label{equation1}
\end{equation}
where $Y_{i,t}$ represents the scalar dependent variable for the $i$th equation at time $t$, and $\mathbf{x}_{i,t} = [x_{1i,t}, x_{2i,t}, \ldots, x_{ki,t}]^{\prime}$ denotes the $k \times 1$ vector of explanatory variables corresponding to the same equation and time period. The model includes a scalar intercept term $\alpha_i$ and a $k \times 1$ vector of slope coefficients $\bm{\beta}_i$, both of which are unknown parameters to be estimated. The stochastic disturbance $e_{i,t}$ is assumed to be unobserved and scalar.

To simplify notation and facilitate subsequent analysis, we introduce the following matrix representations:
\small
\begin{equation}
  \mathbf{Y}_t =
\begin{bmatrix}
 Y_{1,t}\\
 Y_{2,t}\\
 \vdots\\
 Y_{N,t}\\
\end{bmatrix},\
 \mathbf{X}_t=
\begin{bmatrix}
 \mathbf{x}_{1,t}^\prime & \bm{0} & \cdots & \bm{0}\\
 \bm{0} & \mathbf{x}_{2,t}^\prime & \cdots & \bm{0}\\
 \vdots & \vdots & \ddots & \vdots \\
 \bm{0} & \bm{0} & \cdots & \mathbf{x}_{N,t}^\prime\\
\end{bmatrix},\
\bm{\alpha}_0 = 
\begin{bmatrix}
 \alpha_1\\
 \alpha_2\\
 \vdots\\
 \alpha_N\\
\end{bmatrix},\
\bm{\beta}_0 = 
\begin{bmatrix}
 \bm{\beta}_1\\
 \bm{\beta}_2\\
 \vdots\\
 \bm{\beta}_N\\
\end{bmatrix},\
\mathbf{e}_t = 
\begin{bmatrix}
 e_{1,t}\\
 e_{2,t}\\
 \vdots\\
 e_{N,t}\\
\end{bmatrix}.
\label{matrix_notations}
\end{equation}
\normalsize
Here, $\mathbf{X}_t$ is a $N \times K$ matrix, where $K = Nk$. Using this notation, the model in Equation (\ref{equation1}) can be reformulated as follows:
\begin{equation}
\label{vmodel}
\mathbf{Y}_t = \bm{\alpha}_0 + \mathbf{X}_t \bm{\beta}_0 + \mathbf{e}_t
= \mathbf{Z}_t \bm{\kappa}_0+ \mathbf{e}_t,\ \ t=1,...,T,
\end{equation}
where $\mathbf{Z}_t = [\mathbf{I}_K, \mathbf{X}_t]$ is the $N\times (N+K)$ combined design matrix, and $\bm{\kappa}_0=[\bm{\alpha}_0^{\prime}, \bm{\beta}_0^{\prime} ]^{\prime}$ is the $(N+K)\times 1$ vector of parameters. To extend this notation across all time periods, let $\mathbf{Y} = [\mathbf{Y}_1^\prime, \mathbf{Y}_2^\prime, \ldots, \mathbf{Y}_T^\prime]^\prime$ be an $TN \times 1$ vector, $\mathbf{Z} = [\mathbf{Z}_1^\prime, \mathbf{Z}_2^\prime, \ldots, \mathbf{Z}_T^\prime]^\prime$ be an $TN \times (N+K)$ matrix, and $\mathbf{e} = [\mathbf{e}_1^\prime, \mathbf{e}_2^\prime, \ldots, \mathbf{e}_T^\prime]^\prime$ be an $TN \times 1$ vector. Using these definitions, the system can be expressed compactly as:
\begin{equation}
\label{Origmodel}
\mathbf{Y} =
\begin{bmatrix}
 \mathbf{E}_{T,N} & \mathbf{X}
\end{bmatrix}
\begin{bmatrix}
 \bm{\alpha}_0\\
 \bm{\beta}_0\\
\end{bmatrix}
+ \mathbf{e}
= \mathbf{Z}\bm{\kappa}_0+ \mathbf{e},
\end{equation}
where $\mathbf{E}_{T, N} = \bm{\iota}_T \otimes \mathbf{I}_N$, $\bm{\iota}_T$ is a $T \times 1$ vector of ones, and $\mathbf{X} = [\mathbf{X}_1^\prime, \mathbf{X}_2^\prime, \ldots, \mathbf{X}_T^\prime]^\prime$ is a $TN \times K$ matrix, with each component of $\mathbf{X}$ corresponding to $\mathbf{X}_t$ as defined in (\ref{matrix_notations}). 

Furthermore, we assume that the disturbance term $\mathbf{e}_t$ follows a $N$-dimensional VAR($p$) model:
\begin{equation}
\label{VARq}
\mathbf{e}_t = \bm{\Phi}_1\mathbf{e}_{t-1} + \cdots + \bm{\Phi}_p\mathbf{e}_{t-p} + \bm{\varepsilon}_t=\bm{\Phi}\mathbf{V}_t + \bm{\varepsilon}_t,
\end{equation}
where $p$ is an unknown fixed positive integer, $\bm{\Phi}_j$ $(j=1, 2, \ldots, p)$ are $N \times N$ coefficient matrices, and $\bm{\varepsilon}_t$ is a $N \times 1$ white noise vector with a nonsingular covariance matrix $\bm{\Omega}$. Collectively, $\bm{\Phi} = [\bm{\Phi}_1, \bm{\Phi}_2, \ldots, \bm{\Phi}_p]$ forms a $N \times Np$ coefficient matrix, and $\mathbf{V}_t = [\mathbf{e}_{t-1}^\prime, \mathbf{e}_{t-2}^\prime, \ldots, \mathbf{e}_{t-p}^\prime]^\prime$ is a $Np \times 1$ vector.

\begin{assumption}
The coefficient matrices $\bm{\Phi}_j$'s satisfy the stationarity condition, meaning that all solutions $z$ of 
\[
|\mathbf{I}_N - \bm{\Phi}_1z - \bm{\Phi}_2z^2 - \cdots - \bm{\Phi}_pz^p | = 0
\]
lie strictly outside the unit circle (i.e.,$|z|>1$).
\label{assumption1}
\end{assumption}

\begin{assumption}
For $t = 1, \ldots, n$, the random vectors $\bm{\varepsilon}_t$ are independent with $\ex(\bm{\varepsilon}_t) = \bm{0}$ and $\ex(\bm{\varepsilon}_t\bm{\varepsilon}_t^\prime) = \bm{\Omega}$, where $\bm{\Omega}$ is positive definite. Furthermore, for $k_1, k_2, k_3, k_4 = 1, \ldots, N$, the fourth moments satisfy
\[
 \ex|\varepsilon_{k_1,t}\varepsilon_{k_2,t}\varepsilon_{k_3,t}\varepsilon_{k_4,t}|<c
\]
for some finite constant $c$.
\label{assumption2}
\end{assumption}

\begin{assumption}
Suppose the following conditions hold:
\begin{itemize}
 \item[\textnormal{(}i\textnormal{)}] 
$\mathbf{X}_t$ is a stationary and $m$-dependent sequence with $\bm{\mu}_{x,i}\coloneqq \ex(\mathbf{x}_{i,t})$ and $\bm{\Gamma}_{x,ij}^{(s)}\coloneqq \ex(\mathbf{x}_{i,t}\mathbf{x}_{j,t-s}^{\prime})$, where $\bm{\Gamma}_{x,ij}^{(0)}$ is a positive definite matrix. Moreover, $T^{-1}\sum_{t=1}^T \mathbf{x}_{i,t}\mathbf{x}_{j,t}^{\prime}$ is non-singular a.s. for all $T$.
 \item[\textnormal{(}ii\textnormal{)}]
$\varepsilon_{i,t}$ is independent of $x_{j,k,s}$ for all $i$, $j$, $k$, $t$, and $s$. 
\end{itemize}
\label{assumption3}
\end{assumption}

\begin{theorem}
Given Assumptions \ref{assumption1}, \ref{assumption2}, and \ref{assumption3}, the ordinary least squares \textnormal{(OLS)} estimator, $\widehat{\bm{\kappa}}_T^{_{OLS}}$, exists almost surely \textnormal{(a.s.)} for all $T$ sufficiently large, and
\[
\sqrt{T}(\widehat{\bm{\kappa}}_T^{_{OLS}} -\bm{\kappa}_0)\overset{d}{\rightarrow}\mathcal{N}(\bm{0}_{(1+K)N,1},\mathbf{M}_Z^{-1}\bm{\Gamma}_{w}^{\infty}\mathbf{M}_Z^{-1}),
\]
where $\mathbf{M}_Z \coloneqq \ex(\mathbf{Z}^{\prime}\mathbf{Z}/T)$ is positive definite, and $\bm{\Gamma}_{w}^{\infty} \coloneqq \var(T^{-1/2} \mathbf{Z}^{\prime} \bm{e})$ is $O(1)$.
\label{theorem1}
\end{theorem}
\noindent {\bf Proof.} See Online Appendix.

\subsection{The Prais--Winston FGLS Estimator for Multivariate Regression Models}\label{subsec:PW}
In line with \citet{nagakura2024cot}, we derive the Prais--Winston feasible generalized least squares (PW-FGLS) estimator for multivariate regression models. From this point forward, we assume that Assumption \ref{assumption2} holds. Substituting $\mathbf{e}_t = \mathbf{Y}_t - \bm{\alpha} - \mathbf{X}_t\bm{\beta}$ from (\ref{vmodel}) into Equation (\ref{VARq}), the model can be rewritten in its quasi-differenced form as:
\[
\mathbf{Y}_t^{\scriptscriptstyle{QD_2}} = \mathbf{Z}_t^{\scriptscriptstyle{QD_2}} \bm{\kappa}_0 + \bm{\varepsilon}_t ,\quad
\bm{\varepsilon}_t \sim i.i.d.(\mathbf{0}_{N,1}, \bm{\Omega}),\quad t=p+1,...,T,
\]
where $\mathbf{Y}_t^{\scriptscriptstyle{QD_2}}=\mathbf{Y}_t - \sum_{j=1}^p\bm{\Phi}_j\mathbf{Y}_{t-j}$, $\mathbf{Z}_t^{\scriptscriptstyle{QD_2}}=[\mathbf{C}^{\scriptscriptstyle{QD_2}}, \mathbf{X}_t^{\scriptscriptstyle{QD_2}}]$, $\mathbf{X}_t^{\scriptscriptstyle{QD_2}}=\mathbf{X}_t - \sum_{j=1}^p\bm{\Phi}_j \mathbf{X}_{t-j}$ and $\mathbf{C}^{\scriptscriptstyle{QD_2}}=\mathbf{I}_N - \sum_{j=1}^p\bm{\Phi}_j$. By stacking the quasi-differenced variables, we obtain
\begin{equation}
\label{QDmodel2}
\mathbf{Y}^{_{QD_2}} = \mathbf{Z}^{_{QD_2}} \bm{\kappa}_0 + \bm{\varepsilon}^{\scriptscriptstyle{QD_2}},
\end{equation}
where $\mathbf{Y}^{\scriptscriptstyle{QD_2}}=[\mathbf{Y}_{p+1}^{\scriptscriptstyle{QD_2}\prime},\ldots,\mathbf{Y}_T^{\scriptscriptstyle{QD_2}\prime}]^\prime$,\ \ $\mathbf{X}^{\scriptscriptstyle{QD_2}}=[\mathbf{X}_{p+1}^{\scriptscriptstyle{QD_2}\prime}, \ldots, \mathbf{X}_{T}^{\scriptscriptstyle{QD_2}\prime}]^\prime$,\ \ $\mathbf{Z}^{\scriptscriptstyle{QD_2}}=[ \mathbf{E}_{T-p,N}\mathbf{C}^{\scriptscriptstyle{QD_2}}, \mathbf{X}^{\scriptscriptstyle{QD_2}}]$, and $\bm{\varepsilon}^{\scriptscriptstyle{QD_2}}=[\bm{\varepsilon}_{p+1}^\prime,\ldots,\bm{\varepsilon}_{T}^\prime]^\prime$. 

For $t\leq p$, we multiply transformation matrix to Equation (\ref{vmodel}) so that the variance of the error term is equal to $\ve{\Omega}$. In particular, we have the following quasi-differenced model:
\begin{equation}
\label{QDmodel1}
  \mathbf{Y}_t^{\scriptscriptstyle{QD_1}} = \mathbf{Z}_t^{\scriptscriptstyle{QD_1}} \bm{\kappa}_{0} + \bm{\varepsilon}_t^{\scriptscriptstyle{QD_1}},\ \ \bm{\varepsilon}_t^{\scriptscriptstyle{QD_1}} \sim i.i.d.(\mathbf{0}_{N,1},\bm{\Omega}),\quad t=1,...,p,
\end{equation}
where $\mathbf{Y}_t^{\scriptscriptstyle{QD_1}}=\mathbf{\Omega}^{1/2}\bm{\Gamma}_{e}^{\infty-1/2}\mathbf{Y}_t$, $\mathbf{Z}_t^{\scriptscriptstyle{QD_1}}=\mathbf{\Omega}^{1/2}\bm{\Gamma}_{e}^{\infty-1/2}\mathbf{Z}_t$, $\bm{\varepsilon}_t^{\scriptscriptstyle{QD_1}}=\mathbf{\Omega}^{1/2}\bm{\Gamma}_{e}^{\infty-1/2}\mathbf{e}_t$ and $\vecop(\bm{\Gamma}_{e}^{\infty})=(\mathbf{I}_{N^2}-\sum_{j=1}^p(\bm{\Phi}_j\otimes \bm{\Phi}_j))^{-1}\mathrm{vec}(\bm{\Omega})$. Therefore, we have
\begin{equation}
\label{QDmodel1}
\mathbf{Y}^{\scriptscriptstyle{QD_1}} = \mathbf{Z}^{\scriptscriptstyle{QD_1}} \bm{\kappa}_{0} + \bm{\varepsilon}^{\scriptscriptstyle{QD_1}},
\end{equation}
where $\mathbf{Z}^{\scriptscriptstyle{QD_1}}=[ \mathbf{E}_{p,N}\mathbf{C}^{\scriptscriptstyle{QD_1}}, \mathbf{X}^{\scriptscriptstyle{QD_1}}]$, $\mathbf{C}^{\scriptscriptstyle{QD_1}}=\mathbf{\Omega}^{1/2}\bm{\Gamma}_{e}^{\infty-1/2}$, $\mathbf{Y}^{\scriptscriptstyle{QD_1}}=[\mathbf{Y}_{1}^{\scriptscriptstyle{QD_1}\prime}, \mathbf{Y}_{2}^{\scriptscriptstyle{QD_1}\prime}, \ldots, \mathbf{Y}_p^{\scriptscriptstyle{QD_1}\prime}]^\prime$, $\mathbf{X}^{\scriptscriptstyle{QD_1}}=[\mathbf{X}_{1}^{\scriptscriptstyle{QD_1}\prime}, \mathbf{X}_{2}^{\scriptscriptstyle{QD_1}\prime}, \ldots, \mathbf{X}_{p}^{\scriptscriptstyle{QD_1}\prime}]^\prime$, and $\bm{\varepsilon}^{\scriptscriptstyle{QD_1}}=[\bm{\varepsilon}_1^{\scriptscriptstyle{QD_1}\prime}, \bm{\varepsilon}_2^{\scriptscriptstyle{QD_1}\prime}, \ldots, \bm{\varepsilon}_p^{\scriptscriptstyle{QD_1}\prime}]^\prime$. 

By stacking the quasi-differenced variables, we obtain the following model:
\begin{equation}
 \label{SUR_QD}
 \mathbf{Y}^{\scriptscriptstyle{QD}} = \mathbf{Z}^{\scriptscriptstyle{QD}} \bm{\kappa}_0 + \bm{\varepsilon}^{\scriptscriptstyle{QD}},
\end{equation}
where $\mathbf{Y}^{\scriptscriptstyle{QD}}=[\mathbf{Y}^{\scriptscriptstyle{QD_1}\prime},\mathbf{Y}^{\scriptscriptstyle{QD_2}\prime}]^\prime$, $\mathbf{Z}^{\scriptscriptstyle{QD}}=[\mathbf{Z}^{\scriptscriptstyle{QD_1}\prime},\mathbf{Z}^{\scriptscriptstyle{QD_2}\prime}]^\prime$, $\bm{\varepsilon}^{\scriptscriptstyle{QD}}=[\bm{\varepsilon}^{\scriptscriptstyle{QD_1}\prime},\bm{\varepsilon}^{\scriptscriptstyle{QD_2}\prime}]^\prime$. Since $\{\bm{\varepsilon}_t^{\scriptscriptstyle{QD}}\}$ is an i.i.d vector sequence, it follows that $\ex[\bm{\varepsilon}^{\scriptscriptstyle{QD}}\bm{\varepsilon}^{\scriptscriptstyle{QD}\prime}]=\mathbf{I}_{T}\otimes\mathbf{\Omega}$. Replacing $\bm{\Phi}_j^{*} (j=1,...,p)$ and $\bm{\Omega}$ with their consistent esimates, the multivariate PW--FGLS estimator for $\bm{\kappa}_0$ is given by
\begin{equation}
 \label{PW-FGLS}
 \widehat{\bm{\kappa}}_T^{\scriptscriptstyle{PW}}=
\begin{bmatrix}
 \widehat{\bm{\alpha}}_T^{\scriptscriptstyle{PW}}\\
 \widehat{\bm{\beta}}_T^{\scriptscriptstyle{PW}}\\
\end{bmatrix}
=[\widehat{\mathbf{Z}}^{\scriptscriptstyle{QD} \prime}(\mathbf{I}_{T}\otimes \widehat{\bm{\Omega}}^{-1})\widehat{\mathbf{Z}}^{\scriptscriptstyle{QD}}]^{-1}\widehat{\mathbf{Z}}^{\scriptscriptstyle{QD} \prime}(\mathbf{I}_{T}\otimes \widehat{\bm{\Omega}}^{-1})\widehat{\mathbf{Y}}^{\scriptscriptstyle{QD}}.
\end{equation}

The asymptotic properties of the PW-FGLS estimator, $\widehat{\bm{\kappa}}_T^{\scriptscriptstyle{PW}}$, rely on the consistent estimation of $\bm{\Phi}_j (j=1,\ldots,p)$ and $\bm{\Omega}$. To achieve this, we reformulate the VAR($p$) model from (\ref{VARq}) as follows:
\begin{equation}
\label{VAR-lutkepohl}
\mathbf{U} = \bm{\Phi}\mathbf{V} + \mathbf{H},
\end{equation} 
where $\mathbf{U}=[\mathbf{e}_{p+1},\mathbf{e}_{p+2},\ldots,\mathbf{e}_T]$, $\mathbf{V}=[\mathbf{V}_{p+1},\mathbf{V}_{p+2},\ldots,\mathbf{V}_T]$, and $\mathbf{H}=[\bm{\varepsilon}_{p+1},\bm{\varepsilon}_{p+2},\ldots,\bm{\varepsilon}_T]$. Given the lag length $p$, the OLS estimators for $\bm{\Phi}$ and $\bm{\Omega}$ are given by: 
\begin{equation}
 \label{VAR-lutkepohl-ols}
 \widehat{\bm{\Phi}}=[\widehat{\bm{\Phi}}_1,\widehat{\bm{\Phi}}_2,\ldots,\widehat{\bm{\Phi}}_p]=\mathbf{U}\mathbf{V}^\prime(\mathbf{V}\mathbf{V}^\prime)^{-1},\ \textnormal{and}\ \ \widehat{\bm{\Omega}}=\frac{1}{T-p}\mathbf{H}\mathbf{H}^\prime.
\end{equation}
\noindent From Lemma 3 in the Online Appendix, we have $(\mathbf{I}_{N^2}-\sum_{j=1}^p(\widehat{\bm{\Phi}}_j\otimes \widehat{\bm{\Phi}}_j))^{-1}\vecop(\widehat{\bm{\Omega}})\overset{p}{\to}(\mathbf{I}_{N^2}-\sum_{j=1}^p(\bm{\Phi}_j\otimes \bm{\Phi}_j))^{-1}\mathrm{vec}(\bm{\Omega})$, which implies that $\widehat{\bm{\Gamma}}_{e}^{\infty}\overset{p}{\to}\bm{\Gamma}_{e}^{\infty}$. Similarly, we find that $\mathbf{I}_{T}\otimes\widehat{\mathbf{\Omega}}\overset{p}{\to}\mathbf{I}_{T}\otimes\mathbf{\Omega}$. Therefore, the OLS estimator $\widehat{\bm{\Phi}}$ from Equation (\ref{VAR-lutkepohl-ols}) provides consistent estimates of $\bm{\Phi}$, which can then be used to consistently estimate the multivariate PW-FGLS estimator.
\begin{theorem}
%\label{PW-FGLS-asymptotic-distribution}
Suppose that Assumptions \ref{assumption1}, \ref{assumption2} and \ref{assumption3} are satisfied. Then, we have
\begin{equation}
 \label{asymptotic_dist_pw_fgls}
 \sqrt{T}(\widehat{\bm{\kappa}}_T^{\scriptscriptstyle{PW}}-\bm{\kappa}_0) \overset{d}{\rightarrow} \mathcal{N}(\mathbf{0}_{(1+K)N,1},\mathbf{M}_Z^{\scriptscriptstyle{QD_2}-1}),
\end{equation}
where $\mathbf{M}_Z^{\scriptscriptstyle{QD_2}}:=\ex(\mathbf{Z}_t^{\scriptscriptstyle{QD_2}\prime}\bm{\Omega}^{-1}\mathbf{Z}_t^{\scriptscriptstyle{QD_2}})$.
\label{theorem2}
\end{theorem}
\noindent {\bf Proof.} See Online Appendix.

Using only the $QD_2$ variables, we can derive the multivariate CO-FGLS estimator proposed in \citet{nagakura2024cot}, which shares the same asymptotic properties as the multivariate PW-FGLS estimator. Consequently, the multivariate CO-FGLS estimator can be regarded as a special case of the multivariate PW-FGLS estimator.
\begin{corollary}
The multivariate CO--FGLS estimator proposed in \citet{nagakura2024cot} is defined as:
\begin{equation}
 \label{CO-FGLS}
  \widehat{\bm{\kappa}}_T^{\scriptscriptstyle{CO}}=
\begin{bmatrix}
 \widehat{\bm{\alpha}}_T^{\scriptscriptstyle{CO}}\\
 \widehat{\bm{\beta}}_T^{\scriptscriptstyle{CO}}\\
\end{bmatrix}
=
[\widehat{\mathbf{Z}}^{\scriptscriptstyle{QD_2} \prime}(\mathbf{I}_{T-p}\otimes \widehat{\bm{\Omega}}^{-1})\widehat{\mathbf{Z}}^{\scriptscriptstyle{QD_2}}]^{-1}
\widehat{\mathbf{Z}}^{\scriptscriptstyle{QD_2} \prime}(\mathbf{I}_{T-p}\otimes \widehat{\bm{\Omega}}^{-1})\widehat{\mathbf{Y}}^{\scriptscriptstyle{QD_2}},
\end{equation}
and shares the same asymptotic distribution as the multivariate PW--FGLS estimator.
\label{corollary3}
\end{corollary}

\subsection{Wald Tests}\label{subsec:Wald}
In this section, we derive the Wald tests for the null hypothesis $H_0: \mathbf{R}\bm{\kappa}_0 = \mathbf{r}$, using the multivariate PW-FGLS estimator in (\ref{PW-FGLS}) and the multivariate CO-FGLS estimator in (\ref{CO-FGLS}). Specifically, $\mathbf{R}$ is a given $r \times (N + K)$ matrix with $\rank(\mathbf{R}) = r \leq (N + K)$, and $\mathbf{r}$ is a specified $r \times 1$ vector. By leveraging the asymptotic normality of these estimators, we derive the Wald tests to evaluate the null hypothesis $H_0: \mathbf{R}\bm{\kappa}_0 = \mathbf{r}$.
\begin{theorem}
\label{WaldPW}
Suppose that Assumptions \ref{assumption1}, \ref{assumption2}, and \ref{assumption3} hold. Under the null hypothesis $H_0: \mathbf{R}\bm{\kappa}_0=\mathbf{r}$, the Wald test based on the multivariate PW-FGLS estimator is given by:
  \begin{equation}
\label{WaldPW}
\mathcal{W}^{\scriptscriptstyle{PW}}=T(\mathbf{R}\widehat{\bm{\kappa}}_T^{\scriptscriptstyle{PW}}-\mathbf{r})^\prime\left[\mathbf{R}\widehat{\mathbf{M}}_Z^{\scriptscriptstyle{QD}-1}\mathbf{R}^\prime\right]^{-1}(\mathbf{R}\widehat{\bm{\kappa}}_T^{\scriptscriptstyle{PW}}-\mathbf{r})
\overset{d}{\to} \chi_{r}^2,
  \end{equation}
where $\widehat{\mathbf{M}}_Z^{\scriptscriptstyle{QD}}=\sum_{t=1}^T\widehat{\mathbf{Z}}_t^{\scriptscriptstyle{QD}\prime}\widehat{\bm{\Omega}}^{-1}\widehat{\mathbf{Z}}_t^{\scriptscriptstyle{QD}}/T$ and $\chi_r^2$ denotes the chi-squared distribution with $r$ degrees of freedom.
\label{theorem3}
\end{theorem}
\noindent {\bf Proof.} See Online Appendix.

Based on the asymptotic distribution of the multivariate CO-FGLS estimator, the Wald test statistic can be derived as a corollary of the Wald test for the multivariate PW-FGLS estimator under the null hypothesis $H_0: \mathbf{R}\bm{\kappa}_0 = \mathbf{r}$.
\begin{corollary}
\label{WaldCO}
Suppose that Assumptions \ref{assumption1}, \ref{assumption2}, and \ref{assumption3} hold. Under the null hypothesis $H_0: \mathbf{R}\bm{\kappa}_0=\mathbf{r}$, the Wald test based on the multivariate CO-FGLS estimator is defined as:
  \begin{equation}
\label{WaldCO}
\mathcal{W}^{\scriptscriptstyle{CO}}=(T-p)(\mathbf{R}\widehat{\bm{\kappa}}_T^{\scriptscriptstyle{CO}}-\mathbf{r})^\prime\left[\mathbf{R}\widehat{\mathbf{M}}_Z^{\scriptscriptstyle{QD_2}-1}\mathbf{R}^\prime\right]^{-1}(\mathbf{R}\widehat{\bm{\kappa}}_T^{\scriptscriptstyle{CO}}-\mathbf{r})
\overset{d}{\to} \chi_{r}^2,
  \end{equation}
where $\widehat{\mathbf{M}}_Z^{\scriptscriptstyle{QD_2}}=\sum_{t=p+1}^{T}\widehat{\mathbf{Z}}_t^{\scriptscriptstyle{QD}\prime}\widehat{\bm{\Omega}}^{-1}\widehat{\mathbf{Z}}_t^{\scriptscriptstyle{QD}}/(T-p)$ and $\chi_r^2$ represents the chi-squared distribution with $r$ degrees of freedom.
\label{corollary2}
\end{corollary}

In the context of asset pricing models, it is often desirable to evaluate the validity of the models. In such cases, the typical approach is to test the null hypothesis $H_0: \bm{\alpha}_0 = \mathbf{0}_{N,1}$, which can be expressed as:
\[
H_0: \mathbf{R}^{\alpha}\bm{\kappa}_0=\mathbf{0}_{N,1}, \ \ \text{where}\ \ \mathbf{R}^{\alpha}=\left[\mathbf{I}_{N},\ \ \mathbf{0}_{N,K}\right].
\]
Under this null hypothesis, the Wald test statistics can be derived as special cases of Theorem \ref{theorem3} and Corollary \ref{corollary2}.
\begin{corollary}
\label{Wald_alpha}
Suppose that Assumptions \ref{assumption1}, \ref{assumption2}, and \ref{assumption3} hold. Under the null hypothesis $H_0: \bm{\alpha}_0 = \mathbf{0}_{N,1}$, the Wald tests based on the multivariate PW-FGLS and CO-FGLS estimators are defined as follows:
  \begin{align}
\mathcal{W}^{\scriptscriptstyle{PW}}=&T\widehat{\bm{\alpha}}_T^{\scriptscriptstyle{PW}\prime}\left[\mathbf{R}^{\alpha}\widehat{\mathbf{M}}_Z^{\scriptscriptstyle{QD}-1}\mathbf{R}^{\alpha\prime}\right]^{-1}\widehat{\bm{\alpha}}_T^{\scriptscriptstyle{PW}}
\overset{d}{\to} \chi_{N}^2,\label{WaldPW_alpha}\\
\mathcal{W}^{\scriptscriptstyle{CO}}=&(T-p)\widehat{\bm{\alpha}}_T^{\scriptscriptstyle{CO}\prime}\left[\mathbf{R}^{\alpha}\widehat{\mathbf{M}}_Z^{\scriptscriptstyle{QD_2}-1}\mathbf{R}^{\alpha\prime}\right]^{-1}\widehat{\bm{\alpha}}_T^{\scriptscriptstyle{CO}}
\overset{d}{\to} \chi_{N}^2,\label{WaldPW_alpha}
  \end{align}
where $\widehat{\mathbf{M}}_Z^{\scriptscriptstyle{QD}}=\sum_{t=1}^T\widehat{\mathbf{Z}}_t^{\scriptscriptstyle{QD}\prime}\widehat{\bm{\Omega}}^{-1}\widehat{\mathbf{Z}}_t^{\scriptscriptstyle{QD}}/T$, $\widehat{\mathbf{M}}_Z^{\scriptscriptstyle{QD_2}}=\sum_{t=1}^T\widehat{\mathbf{Z}}_t^{\scriptscriptstyle{QD_2}\prime}\widehat{\bm{\Omega}}^{-1}\widehat{\mathbf{Z}}_t^{\scriptscriptstyle{QD_2}}/(T-p)$, and $\chi_N^2$ represents the chi-squared distribution with $N$ degrees of freedom.
\label{corollary3}
\end{corollary}

\section{Simulation Experiment}\label{sec:fgls_grs_sim}

In this section, we conduct a simulation experiment to evaluate the finite-sample performance of the proposed Wald tests. Specifically, we compare them with the original and modified GRS tests and the HAR Wald test under the null hypothesis $H_0: \bm{\alpha}_0=\mathbf{0}_{N,1}$, which is commonly used to assess asset pricing models.

\subsection{Comparative Models}
\citet{gibbons1989teg} originally define the GRS test statistic as
\begin{equation}
\label{GRS}
GRS = \frac{T(T-N-L)}{N(T-L-1)}
(1+\bar{\mathbf{x}}^{\prime}\mathbf{S}_x^{-1}\bar{\mathbf{x}})^{-1}\widehat{\bm{\alpha}}_{T}^{\scriptstyle{OLS}\prime}\widehat{\bm{\Sigma}}^{-1}\widehat{\bm{\alpha}}_{T}^{\scriptstyle{OLS}},\nonumber
\end{equation}
where $T$, $N$, and $L$ denote the number of time-series observations, portfolios, and risk factors, respectively. The other variables are defined as follows:
\begin{equation}
\label{GRSdef}
\bar{\mathbf{x}}=\frac{1}{T}\sum_{t=1}^T \mathbf{x}_t,\ \mathbf{S}_x = \frac{1}{T-1}\sum_{t=1}^T (\mathbf{x}_t -\bar{\mathbf{x}})(\mathbf{x}_t -\bar{\mathbf{x}})^{\prime},\ \widehat{\bm{\Sigma}} = \frac{1}{T-L-1} \sum_{t=1}^T \widehat{\mathbf{e}}_t \widehat{\mathbf{e}}_t^{\prime}.\nonumber
\end{equation}
As pointed out by \citet{kamstra2024tra}, \citet{gibbons1989teg} somewhat loosely define the sample variance-covariance matrix $\mathbf{S}_x$. Consequently, the GRS test statistic does not exactly follow the $F(N, T-N-L)$ distribution under the null hypothesis and tends to over-reject it. To address this issue, \citet{kamstra2024tra} propose a modified GRS test statistic:
\begin{equation}
\label{CGRS}
GRS^{KS} = \frac{T(T-N-L)}{N(T-L-1)}
(1+\bar{\mathbf{x}}^{\prime}\mathbf{S}_x^{\ast-1}\bar{\mathbf{x}})^{-1}\widehat{\bm{\alpha}}_{T}^{\scriptstyle{OLS}\prime}\widehat{\bm{\Sigma}}^{-1}\widehat{\bm{\alpha}}_{T}^{\scriptstyle{OLS}},\nonumber
\end{equation}
where $\mathbf{S}_x^{\ast}$ is defined as $\mathbf{S}_x^{\ast}=\frac{T-1}{T} \mathbf{S}_x$, which directly implies that
\[
GRS =GRS^{KS} + O_p(T^{-1}).
\]
Therefore, $GRS$ and $GRS^{KS}$ are expected to take very similar values when $T$ is relatively large.

The original and modified GRS tests are among the most widely used tests for evaluating the validity of asset pricing models. However, as suggested by \citet{affleck1989nta} and \citet{zhou1993apt}, these tests may perform poorly when the distribution of error terms deviates significantly from normality. To address this issue, \citet{kiefer2000srt} and \citet{kiefer2002hars} propose the HAR test, which accounts for such deviations by extending the HAC estimator introduced by \citet{newey1987sps, newey1994als}. Under the null hypothesis $H_0:\bm{\alpha}_0=\mathbf{0}_{N,1}$, the HAR Wald test statistic is given by:
\begin{equation}
\label{WaldHAR}
 \mathcal{W}^{\scriptstyle{HAR}}=T\widehat{\bm{\alpha}}_{T}^{\scriptstyle{OLS}\prime}\left[\mathbf{R}(\widehat{\mathbf{M}}_Z^{-1}\widehat{\mathbf{\Gamma}}_w^{\infty}\widehat{\mathbf{M}}_Z^{-1})^{-1}\mathbf{R}^\prime\right]^{-1}\widehat{\bm{\alpha}}_{T}^{\scriptstyle{OLS}},\nonumber
\end{equation}
where
\begin{equation}
\begin{array}{l}
\widehat{\mathbf{\Gamma}}_w^{\infty}=\widehat{\mathbf{\Gamma}}_{w,0}+\sum_{j=1}^l w(j,l)\left[\widehat{\mathbf{\Gamma}}_{w,j}+\widehat{\mathbf{\Gamma}}_{w,j}^\prime\right],\ w(j,l)=1-\left[\frac{j}{l+1}\right],\
\widehat{\mathbf{\Gamma}}_{w,j}=\frac{1}{T}\sum_{j+1}^T\widehat{\mathbf{w}}_t\widehat{\mathbf{w}}_{t-j}^\prime,\\ \widehat{\mathbf{w}}_t=\mathbf{Z}_t^\prime\widehat{\mathbf{e}}_t,\ \mbox{and}\  \mathbf{R}=[\mathbf{I}_N,\ \mathbf{0}_{N,KN}].
\end{array}\nonumber
\end{equation}
The lag truncation number $l$ is set to $[4(T/100)]^{(2/9)}$, following \citet{newey1994als}. Under the null hypothesis, the HAR Wald test statistic asymptotically follows a chi-square distribution with $N$ degrees of freedom.

\subsection{Simulation Design}
In asset pricing models, it is common for factors to take the same values across assets. For example, \citeapos{fama1993crf}{fama1993crf,fama2015ffa} multifactor models and the arbitrage pricing theory introduced by \citet{ross1976atc} are typical examples of such models. Accordingly, in our simulation experiments, we focus on scenarios where $\mathbf{x}_{i,t}$ is identical across $i$. Specifically, we set the parameters based on the works of \citet{kiefer2002hars} and \citet{phillips2007lrv}:
\begin{equation}
\label{Setting}
\begin{array}{l}
\mathbf{Y}_t =\bm{\alpha}_0 + \mathbf{X}_t \bm{\beta}_0 + \mathbf{e}_t,\
\mathbf{X}_t = \mathbf{I}_N \otimes \mathbf{x}_{t}^\prime,\
\mathbf{x}_t = 0.5\mathbf{x}_{t-1} + \bm{\eta}_t,\
\bm{\eta}_t \sim\mathcal{N}(\mathbf{0}_{k,1},\ \mathbf{I}_k),\
\bm{\beta}= \bm{\iota}_{Nk},\\
\mathbf{e}_t = \bm{\Phi}_{1} \mathbf{e}_{t-1} +\mathbf{u}_t,\
\mathbf{u}_t \sim\mathcal{N}(\mathbf{0}_{N,1}, \bm{\Omega}),\
\mathbf{x}_0=\mathbf{e}_0=\mathbf{0},\
\mbox{and}\ \cov(\mathbf{u}_t,\bm{\eta}_t)=\mathbf{0},
\end{array}
\nonumber
\end{equation}
where 
\small
\[
 \bm{\Omega}_{ij}=
\begin{cases}
 \sigma_i^2, & \text{if}\ i=j,\\
 \rho\sigma_i\sigma_j, & \text{if}\ i\neq j.
\end{cases}
\]
\normalsize
By specifying $\bm{\Omega}$ as above, we can accommodate not only heteroskedasticity but also cross-sectional dependence. In our simulation, $\sigma_i^2$ is drawn from the uniform distribution $U(0.5,1)$, and $\rho$ is set to $0.3$.

The intercept $\bm{\alpha}_0 = [\alpha_1, ..., \alpha_N]^{\prime}$ is set to $\bm{\alpha}_0 = \mathbf{0}_{N,1}$ under the null hypothesis, while under the alternative hypothesis, it is specified as $\alpha_1 = 0.1$ and $\alpha_j = 0$ for $j = 2,\ldots, N$. We consider the following values for $T$, $N$, and $k$: $T \in \{200, 400, 800, 1600, 3200\}$; $k \in \{3, 5\}$; and $N \in \{6, 25\}$. The number of iterations is set to 1000. We examine two cases for the error terms: (i) heteroskedastic error terms and (ii) heteroskedastic and autocorrelated error terms. In the latter case, we assume that the autocorrelated errors follow a VAR(1) process. We then set $\bm{\Phi}_{1} = \mathbf{0}$ for case (i) and $\bm{\Phi}_{1} = 0.3\times \mathbf{I}_N$ for case (ii).\footnote{In each case, we use the Bayesian information criterion (BIC) of \citet{schwarz1978edm} to determine the optimal lag length of the VAR model.}

\subsection{Simulation Results}
In the following, we present the simulation results based on the above settings. Specifically, we assess whether the empirical rejection rates of the test statistics align with the nominal significance levels under the null hypothesis, which corresponds to evaluating the Type I error rate.\footnote{We confirm that the power of the tests (i.e., the rejection rate under the alternative hypothesis) also perform well (see Tables A.1 and A.2 in the \href{https://at-noda.com/appendix/fgls_grs_appendix.pdf}{Online Appendix}).}

\begin{center}
(Table \ref{fgls_grs_tab1} around here)
\end{center}

Table \ref{fgls_grs_tab1} presents the rejection rates (Type I error rates) of the test statistics under the null hypothesis with heteroskedastic errors. The results indicate that the empirical sizes of these test statistics are generally close to the nominal significance levels, regardless of the test type.

\begin{center}
(Table \ref{fgls_grs_tab2} around here)
\end{center}

Table \ref{fgls_grs_tab2} reports the rejection frequencies of the test statistics under the null hypothesis with heteroskedastic and autocorrelated errors. The results show that our proposed Wald tests exhibit accurate size properties, with empirical sizes converging to the nominal levels as the sample size increases, regardless of the number of portfolios.

In contrast, the other test statistics perform poorly. Both the original and modified GRS tests exhibit substantial size distortions relative to the normal error case. Our findings confirm that this issue persists regardless of $N$ when the assumption of independent errors is violated. Furthermore, while the HAR Wald test performs well for $N = 6$, it severely over-rejects the null hypothesis for $N = 25$, indicating reduced reliability as the number of portfolios increases.

Overall, we confirm that the proposed Wald test performs well in the presence of both heteroskedasticity and autocorrelation. In contrast, the original and modified GRS tests, as well as the HAR test, exhibit a severe over-rejection tendency when the error terms are heteroskedastic and autocorrelated. Therefore, we conclude that the proposed Wald statistics have desirable properties even when the assumptions about the error terms are relaxed.

\section{Conclusion}\label{sec:fgls_grs_con}
In this paper, we study the asymptotic properties of the GLS estimator in multivariate regression with heteroskedastic and autocorrelated errors. We also constuct Wald statistics for linear restrictions on the estimator and assess their finite-sample performance. As a result of the simulation experiment, we find that our Wald test performs well regardless of the presence of heteroskedasticity and autocorrelation, differing from existing test statistics.

\section*{Acknowledgments}
The authors would like to thank Tirthatanmoy Das, Akitada Kasahara, Keiichi Morimoto, Daisuke Nagakura, Tatsuma Wada, Yohei Yamamoto, Taiyo Yoshimi, the conference participants at the Japan Society of Monetary Economics 2024 Autumn Meeting and the Western Economic Association International 99th Annual Conference for their helpful comments and suggestions. The author (Noda) is also grateful for the financial assistance provided by the Japan Society for the Promotion of Science Grant in Aid for Scientific Research (grant number: 23H00838) and Japan Science and Technology Agency, Moonshot Research \& Development Program (grant number: JPMJMS2215). All data and programs used are available upon request.

%\clearpage

\clearpage

\bigskip

\bigskip

\bigskip

\setcounter{table}{0}
\renewcommand{\thetable}{\arabic{table}}

%\begin{landscape}
\begin{table}[p]
\caption{Rejection rates under the null hypothesis $H_0:\bm{\alpha}_0=\mathbf{0}_{N,1}$ (heteroskedastic errors)}\label{fgls_grs_tab1}
\begin{center}\resizebox{16cm}{!}{
\begin{tabular}{lllccccccccccccccccccccc} \hline\hline
 &  &  &  & \multicolumn{3}{c}{$\mathcal{W}^{\scriptstyle{PW}}$} & & \multicolumn{3}{c}{$\mathcal{W}^{\scriptstyle{CO}}$} & & \multicolumn{3}{c}{$\mathcal{W}^{\scriptstyle{HAR}}$} & & \multicolumn{3}{c}{GRS} & & \multicolumn{3}{c}{GRS$^{\scriptstyle{KS}}$} \\\cline{5-23}
 &  &  &  & 10\% & 5\% & 1\% &  & 10\% & 5\% & 1\% &  & 10\% & 5\% & 1\% &  & 10\% & 5\% & 1\% &  & 10\% & 5\% & 1\% & \\\cline{5-7}\cline{9-11}\cline{13-15}\cline{17-19}\cline{21-23}
 & $N=6/K=3$ &  &  &  &  &  &  &  &  &  &  &  &  &  &  &  &  &  &  &  &  &  &  \\
 &  & $T=200$ &  & $0.163$ & $0.096$ & $0.029$ &  & $0.160$ & $0.092$ & $0.026$ &  & $0.177$ & $0.119$ & $0.033$ &  & $0.098$ & $0.046$ & $0.008$ &  & $0.098$ & $0.046$ & $0.008$ & \\
 &  & $T=400$ &  & $0.110$ & $0.062$ & $0.014$ &  & $0.108$ & $0.064$ & $0.012$ &  & $0.135$ & $0.068$ & $0.021$ &  & $0.087$ & $0.043$ & $0.007$ &  & $0.087$ & $0.043$ & $0.007$ & \\
 &  & $T=800$ &  & $0.112$ & $0.051$ & $0.007$ &  & $0.115$ & $0.052$ & $0.008$ &  & $0.124$ & $0.071$ & $0.007$ &  & $0.100$ & $0.047$ & $0.008$ &  & $0.100$ & $0.047$ & $0.008$ & \\
 &  & $T=1600$ &  & $0.114$ & $0.055$ & $0.013$ &  & $0.117$ & $0.057$ & $0.013$ &  & $0.120$ & $0.058$ & $0.014$ &  & $0.109$ & $0.047$ & $0.011$ &  & $0.109$ & $0.047$ & $0.011$ & \\
 &  & $T=3200$ &  & $0.105$ & $0.043$ & $0.011$ &  & $0.108$ & $0.044$ & $0.011$ &  & $0.108$ & $0.047$ & $0.012$ &  & $0.102$ & $0.036$ & $0.012$ &  & $0.102$ & $0.036$ & $0.012$ & \\\hdashline
 & $N=6/K=5$ &  &  &  &  &  &  &  &  &  &  &  &  &  &  &  &  &  &  &  &  &  &  \\
 &  & $T=200$ &  & $0.175$ & $0.101$ & $0.024$ &  & $0.168$ & $0.104$ & $0.024$ &  & $0.193$ & $0.129$ & $0.032$ &  & $0.089$ & $0.043$ & $0.007$ &  & $0.089$ & $0.043$ & $0.007$ & \\
 &  & $T=400$ &  & $0.126$ & $0.070$ & $0.023$ &  & $0.129$ & $0.071$ & $0.025$ &  & $0.145$ & $0.085$ & $0.028$ &  & $0.091$ & $0.056$ & $0.008$ &  & $0.091$ & $0.056$ & $0.008$ & \\
 &  & $T=800$ &  & $0.105$ & $0.063$ & $0.009$ &  & $0.107$ & $0.061$ & $0.009$ &  & $0.112$ & $0.068$ & $0.014$ &  & $0.087$ & $0.053$ & $0.008$ &  & $0.087$ & $0.053$ & $0.008$ & \\
 &  & $T=1600$ &  & $0.088$ & $0.054$ & $0.014$ &  & $0.089$ & $0.053$ & $0.012$ &  & $0.098$ & $0.055$ & $0.013$ &  & $0.084$ & $0.046$ & $0.011$ &  & $0.084$ & $0.046$ & $0.011$ & \\
 &  & $T=3200$ &  & $0.093$ & $0.053$ & $0.011$ &  & $0.093$ & $0.054$ & $0.011$ &  & $0.095$ & $0.053$ & $0.009$ &  & $0.088$ & $0.049$ & $0.009$ &  & $0.088$ & $0.049$ & $0.009$ & \\\hdashline
 & $N=25/K=3$ &  &  &  &  &  &  &  &  &  &  &  &  &  &  &  &  &  &  &  &  &  \\
 &  & $T=200$ &  & $0.632$ & $0.535$ & $0.348$ &  & $0.630$ & $0.539$ & $0.347$ &  & $0.745$ & $0.663$ & $0.491$ &  & $0.111$ & $0.054$ & $0.008$ &  & $0.111$ & $0.054$ & $0.008$ & \\
 &  & $T=400$ &  & $0.327$ & $0.238$ & $0.098$ &  & $0.327$ & $0.232$ & $0.096$ &  & $0.457$ & $0.351$ & $0.177$ &  & $0.099$ & $0.042$ & $0.007$ &  & $0.099$ & $0.042$ & $0.007$ & \\
 &  & $T=800$ &  & $0.201$ & $0.122$ & $0.027$ &  & $0.198$ & $0.122$ & $0.028$ &  & $0.274$ & $0.188$ & $0.060$ &  & $0.093$ & $0.040$ & $0.004$ &  & $0.093$ & $0.040$ & $0.004$ & \\
 &  & $T=1600$ &  & $0.134$ & $0.070$ & $0.014$ &  & $0.134$ & $0.068$ & $0.015$ &  & $0.187$ & $0.100$ & $0.023$ &  & $0.078$ & $0.043$ & $0.008$ &  & $0.078$ & $0.043$ & $0.008$ & \\
 &  & $T=3200$ &  & $0.121$ & $0.065$ & $0.009$ &  & $0.122$ & $0.066$ & $0.009$ &  & $0.144$ & $0.082$ & $0.015$ &  & $0.095$ & $0.048$ & $0.005$ &  & $0.095$ & $0.048$ & $0.005$ & \\\hdashline
 & $N=25/K=5$ &  &  &  &  &  &  &  &  &  &  &  &  &  &  &  &  &  &  &  &  &  \\
 &  & $T=200$ &  & $0.661$ & $0.547$ & $0.377$ &  & $0.651$ & $0.553$ & $0.372$ &  & $0.784$ & $0.690$ & $0.549$ &  & $0.120$ & $0.064$ & $0.010$ &  & $0.120$ & $0.064$ & $0.010$ & \\
 &  & $T=400$ &  & $0.328$ & $0.239$ & $0.111$ &  & $0.330$ & $0.232$ & $0.109$ &  & $0.492$ & $0.360$ & $0.208$ &  & $0.100$ & $0.044$ & $0.004$ &  & $0.100$ & $0.044$ & $0.004$ & \\
 &  & $T=800$ &  & $0.205$ & $0.120$ & $0.032$ &  & $0.203$ & $0.117$ & $0.032$ &  & $0.290$ & $0.187$ & $0.068$ &  & $0.090$ & $0.047$ & $0.004$ &  & $0.090$ & $0.047$ & $0.004$ & \\
 &  & $T=1600$ &  & $0.139$ & $0.071$ & $0.019$ &  & $0.139$ & $0.071$ & $0.019$ &  & $0.178$ & $0.107$ & $0.025$ &  & $0.096$ & $0.049$ & $0.010$ &  & $0.096$ & $0.049$ & $0.010$ & \\
 &  & $T=3200$ &  & $0.121$ & $0.068$ & $0.013$ &  & $0.121$ & $0.068$ & $0.012$ &  & $0.153$ & $0.086$ & $0.023$ &  & $0.100$ & $0.057$ & $0.009$ &  & $0.100$ & $0.057$ & $0.009$ & \\\hline\hline
\end{tabular}}
{\resizebox{16cm}{!}{\begin{minipage}{750pt}
\vspace{3mm}
{\underline{Note:}} R version 4.4.2 was used to compute the statistics.
\end{minipage}}}
\end{center}
\end{table}
%\end{landscape}

%\clearpage

%\begin{landscape}
\begin{table}[p]
\caption{Rejection rates under the null hypothesis $H_0:\bm{\alpha}_0=\mathbf{0}_{N,1}$ (heteroskedastic and autocorrelated errors)}\label{fgls_grs_tab2}
\begin{center}\resizebox{16cm}{!}{
\begin{tabular}{lllccccccccccccccccccccc} \hline\hline
 &  &  &  & \multicolumn{3}{c}{$\mathcal{W}^{\scriptstyle{PW}}$} & & \multicolumn{3}{c}{$\mathcal{W}^{\scriptstyle{CO}}$} & & \multicolumn{3}{c}{$\mathcal{W}^{\scriptstyle{HAR}}$} & & \multicolumn{3}{c}{GRS} & & \multicolumn{3}{c}{GRS$^{\scriptstyle{KS}}$} \\\cline{5-23}
 &  &  &  & 10\% & 5\% & 1\% &  & 10\% & 5\% & 1\% &  & 10\% & 5\% & 1\% &  & 10\% & 5\% & 1\% &  & 10\% & 5\% & 1\% & \\\cline{5-7}\cline{9-11}\cline{13-15}\cline{17-19}\cline{21-23}
 & $N=6/K=3$ &  &  &  &  &  &  &  &  &  &  &  &  &  &  &  &  &  &  &  &  &  &  \\
 &  & $T=200$ &  & $0.195$ & $0.127$ & $0.046$ &  & $0.192$ & $0.122$ & $0.045$ &  & $0.258$ & $0.176$ & $0.069$ &  & $0.436$ & $0.324$ & $0.154$ &  & $0.435$ & $0.324$ & $0.154$ & \\
 &  & $T=400$ &  & $0.126$ & $0.076$ & $0.023$ &  & $0.127$ & $0.068$ & $0.022$ &  & $0.197$ & $0.114$ & $0.043$ &  & $0.432$ & $0.335$ & $0.150$ &  & $0.431$ & $0.335$ & $0.150$ & \\
 &  & $T=800$ &  & $0.119$ & $0.058$ & $0.009$ &  & $0.122$ & $0.060$ & $0.009$ &  & $0.171$ & $0.097$ & $0.024$ &  & $0.432$ & $0.325$ & $0.168$ &  & $0.432$ & $0.325$ & $0.168$ & \\
 &  & $T=1600$ &  & $0.122$ & $0.058$ & $0.014$ &  & $0.120$ & $0.061$ & $0.014$ &  & $0.159$ & $0.089$ & $0.022$ &  & $0.455$ & $0.336$ & $0.175$ &  & $0.455$ & $0.336$ & $0.175$ & \\
 &  & $T=3200$ &  & $0.109$ & $0.046$ & $0.011$ &  & $0.110$ & $0.047$ & $0.012$ &  & $0.141$ & $0.075$ & $0.018$ &  & $0.443$ & $0.324$ & $0.160$ &  & $0.443$ & $0.324$ & $0.160$ & \\\hdashline
 & $N=6/K=5$ &  &  &  &  &  &  &  &  &  &  &  &  &  &  &  &  &  &  &  &  &  & \\
 &  & $T=200$ &  & $0.220$ & $0.138$ & $0.042$ &  & $0.217$ & $0.137$ & $0.042$ &  & $0.293$ & $0.205$ & $0.082$ &  & $0.435$ & $0.324$ & $0.154$ &  & $0.435$ & $0.324$ & $0.153$ & \\
 &  & $T=400$ &  & $0.149$ & $0.087$ & $0.031$ &  & $0.159$ & $0.087$ & $0.028$ &  & $0.212$ & $0.133$ & $0.042$ &  & $0.436$ & $0.310$ & $0.150$ &  & $0.436$ & $0.309$ & $0.150$ & \\
 &  & $T=800$ &  & $0.116$ & $0.066$ & $0.012$ &  & $0.119$ & $0.069$ & $0.014$ &  & $0.158$ & $0.094$ & $0.033$ &  & $0.431$ & $0.327$ & $0.153$ &  & $0.431$ & $0.327$ & $0.153$ & \\
 &  & $T=1600$ &  & $0.096$ & $0.055$ & $0.014$ &  & $0.094$ & $0.053$ & $0.013$ &  & $0.140$ & $0.078$ & $0.022$ &  & $0.416$ & $0.307$ & $0.142$ &  & $0.416$ & $0.307$ & $0.142$ & \\
 &  & $T=3200$ &  & $0.094$ & $0.055$ & $0.011$ &  & $0.091$ & $0.054$ & $0.012$ &  & $0.126$ & $0.073$ & $0.017$ &  & $0.426$ & $0.302$ & $0.156$ &  & $0.426$ & $0.302$ & $0.156$ & \\\hdashline
 & $N=25/K=3$ &  &  &  &  &  &  &  &  &  &  &  &  &  &  &  &  &  &  &  &  &  & \\
 &  & $T=200$ &  & $0.815$ & $0.741$ & $0.583$ &  & $0.815$ & $0.745$ & $0.586$ &  & $0.893$ & $0.848$ & $0.736$ &  & $0.807$ & $0.717$ & $0.484$ &  & $0.807$ & $0.717$ & $0.483$ & \\
 &  & $T=400$ &  & $0.486$ & $0.367$ & $0.188$ &  & $0.483$ & $0.365$ & $0.197$ &  & $0.666$ & $0.547$ & $0.355$ &  & $0.797$ & $0.711$ & $0.516$ &  & $0.797$ & $0.711$ & $0.516$ & \\
 &  & $T=800$ &  & $0.274$ & $0.175$ & $0.054$ &  & $0.274$ & $0.180$ & $0.056$ &  & $0.434$ & $0.320$ & $0.156$ &  & $0.817$ & $0.724$ & $0.500$ &  & $0.817$ & $0.724$ & $0.500$ & \\
 &  & $T=1600$ &  & $0.159$ & $0.095$ & $0.022$ &  & $0.159$ & $0.094$ & $0.024$ &  & $0.279$ & $0.191$ & $0.063$ &  & $0.817$ & $0.731$ & $0.524$ &  & $0.817$ & $0.731$ & $0.524$ & \\
 &  & $T=3200$ &  & $0.126$ & $0.073$ & $0.010$ &  & $0.128$ & $0.073$ & $0.010$ &  & $0.246$ & $0.140$ & $0.044$ &  & $0.802$ & $0.720$ & $0.527$ &  & $0.802$ & $0.720$ & $0.527$ & \\\hdashline
 & $N=25/K=5$ &  &  &  &  &  &  &  &  &  &  &  &  &  &  &  &  &  &  &  &  &  & \\
 &  & $T=200$ &  & $0.847$ & $0.766$ & $0.608$ &  & $0.844$ & $0.770$ & $0.607$ &  & $0.915$ & $0.877$ & $0.768$ &  & $0.798$ & $0.706$ & $0.467$ &  & $0.798$ & $0.703$ & $0.467$ & \\
 &  & $T=400$ &  & $0.507$ & $0.388$ & $0.201$ &  & $0.503$ & $0.382$ & $0.200$ &  & $0.668$ & $0.571$ & $0.385$ &  & $0.790$ & $0.708$ & $0.489$ &  & $0.790$ & $0.708$ & $0.489$ & \\
 &  & $T=800$ &  & $0.266$ & $0.180$ & $0.057$ &  & $0.267$ & $0.179$ & $0.057$ &  & $0.449$ & $0.330$ & $0.156$ &  & $0.807$ & $0.717$ & $0.516$ &  & $0.807$ & $0.717$ & $0.516$ & \\
 &  & $T=1600$ &  & $0.162$ & $0.096$ & $0.024$ &  & $0.166$ & $0.095$ & $0.025$ &  & $0.297$ & $0.186$ & $0.074$ &  & $0.831$ & $0.715$ & $0.512$ &  & $0.831$ & $0.715$ & $0.512$ & \\
 &  & $T=3200$ &  & $0.143$ & $0.077$ & $0.018$ &  & $0.144$ & $0.079$ & $0.019$ &  & $0.240$ & $0.147$ & $0.047$ &  & $0.803$ & $0.732$ & $0.541$ &  & $0.803$ & $0.732$ & $0.541$ & \\\hline\hline
\end{tabular}}
{\resizebox{16cm}{!}{\begin{minipage}{750pt}
\vspace{3mm}
{\underline{Note:}} R version 4.4.2 was used to compute the statistics.
\end{minipage}}}
\end{center}
\end{table}
%\end{landscape}

\clearpage

\end{document}